\documentstyle[aasms4,psfig]{article}
\newcommand{\be}{\begin{equation}}
\newcommand{\ba}{\begin{eqnarray}}
\newcommand{\ee}{\end{equation}}
\newcommand{\ea}{\end{eqnarray}}

\newcommand{\etal}{{\it et al. }}
\lefthead{Nakamura \& Chiba }
\righthead{}
\begin{document}
\title{Determining  the Equation of State of the Expanding Universe
  Using a New Independent Variable }
\author{Takashi Nakamura}
\affil{Yukawa Institute for Theoretical Physics, Kyoto University,
 Kyoto 606-8502}
\author{Takeshi Chiba}
\affil{Department of Physics, Kyoto University, Kyoto 606-8502}

%\received{2000 August 31 }
%\accepted{2000 August 31}

\begin{abstract}
To determine the equation of state of the universe, we propose to use a
new independent variable  $R\equiv (H_0/c)(d_L(z)/(1+z))$, where $H_0$
and $d_L(z)$ are the present Hubble parameter and the luminosity
distance, respectively. For the flat universe suggested from the
observation of the anisotropy of cosmic microwave background, the
density and the pressure are expressed as
$\rho/\rho_0=4(df/dR)^2/f^6$ and $p/\rho_0=-4/3(d^2f/dR^2)/f^5$ where 
$\rho_0$ is the present density and $f(R)=1/\sqrt{1+z(R)}$. 
In $(R, f)$ plane the sign as well as the strength of the pressure is in 
proportion to the curvature of the curve $f(R)$. We propose to adopt a 
Pade-like expression of $f(R)=1/\sqrt{u}$ with 
$u\equiv 1+\sum\limits_{n=1}^{N}u_nR^n$. For flat $\Lambda$ model the 
expansion up to $N=7$ has at most an error $< 0.2\%$ for $z < 1.7$ and
any value of $\Lambda$. We also propose a general method to determine
the equation of state of the universe which has $N-1$ free parameters. 
If the number of parameters are smaller than $N-1$, there is a
consistency check of the equation of state so that we may confirm or
refute each model.
\end{abstract}


\section{Introduction}

Recent measurements of the luminosity distance $d_L(z)$ using
Type Ia supernovae (\cite{schmidt98,ries98,perl99}) suggest that 
accurate $d_L(z)$ may be obtained in the near future. Especially SNAP 
(\cite{snap}) will give us the luminosity distance of $\sim$2000 Type Ia 
supernovae with an accuracy of a few \% up to $z\sim1.7$ every year.  
On the other hand from the observation of the first Doppler peak of
the anisotropy of CMB,  it is now suggested that the universe is
flat (\cite{ber00,lange00}), which may be proved  in future by MAP and
Planck. If the flat universe is the case, the density $\rho(z)$ and the 
pressure $p(z)$ can be determined only from  $d_L(z)$  in principle 
(\cite{naka99}) so that the equation of the state of the universe is 
uniquely determined. If not, the determination of the present curvature
of the universe  and the determination of $\rho(z)$ and  $p(z)$ will be 
coupled in general (\cite{naka99}).

Now let us assume that the universe is flat.
Even in this case at least two problems exist: 1) How to express the
continuous function $d_L(z)$ which is  accurate enough
from $z=0$ to $z\sim 1.7$ using several free parameters; 2) From
$d_L(z)$ how to obtain accurate  $\rho(z)$ and  $p(z)$, that is, 
the equation of state of the 
universe(\cite{star98,huter99,naka99,saini99,chiba00}). 
        
In this paper, we  propose 
to use a new independent variable  $R\equiv (H_0/c)(d_L(z)/(1+z))$
instead of $z$, where $H_0$ is the present Hubble
parameter. We show that  $\rho(R)/\rho_0=4(df/dR)^2/f^6$ and
$p(R)/\rho_0=-4/3(d^2f/dR^2)/f^5$ where $\rho_0$ is the present density
and $f(R)=1/\sqrt{1+z(R)}$. This means that the pressure is in
proportion to the curvature of the curve $f(R)$. For an accurate
expression of $f(R)$  we propose  a Pade-like form of $f(R)=1/\sqrt{u}$ 
with $u\equiv 1+\sum\limits_{n=1}^{N} u_nR^n$.
For flat $\Lambda$ model, the expansion up to $N=7$ has at most an
error $< 0.2\%$  for $z < 1.7$ and all value of $\Lambda$.
We also propose a general method    to determine 
the equation of state of the universe which has  $< N-1$ free parameters.

\section{New Variables}

The luminosity distance $d_L(z)$ is given by 
$ d_L(z)=a_0(1+z)f(\chi)$, with
 $\chi=1/ a_0\int^z_0{dz'/ H(z')}$,
where $H(z)$ and $a_0$ are the Hubble parameter at $z$ and the
present scale factor,  and  $f(\chi)=\chi, \sinh (\chi)$ and $\sin
(\chi)$ for flat, open and closed universe,
 respectively. Let  us define $R=(H_0/c)(d_L(z)/(1+z))$, 
where $H_0$ is the present Hubble parameter. 
Then  $\rho(z)$ and $p(z)$ are expressed as 
\ba
&&\frac{\rho(z)}{\rho_0}=  \frac{1}{(\frac{dR}{dz})^2}+\left((1+z)^2-
\frac{R^2}{(\frac{dR}{dz})^2}\right)H_0^2 \Omega_{k0},\nonumber \\
&&\frac{3p(z)}{\rho_0}=  - \frac{3}{(\frac{dR}{dz})^2}+
(1+z)\frac{d}{dz}\left(\frac{1}{(\frac{dR}{dz})^2}\right)
- \left((1+z)^2-\frac{3R^2}{(\frac{dR}{dz})^2}+(1+z)\frac{d}{dz}
\left(\frac{R^2}
{(\frac{dR}{dz})^2}\right)\right)H_0^2\Omega_{k0},
\ea
where  $\rho_0$ is the present density and $\Omega_{k0}\equiv
k/(a_0^2H_0^2)$ (\cite{naka99}).
Since the flat universe is suggested both from observations
(\cite{ber00,lange00}) and theory 
(inflation paradigm), we consider only  $\Omega_{k0}=0$ case in this
paper.

Now we adopt $R$ as an independent variable instead of $z$. Then
$\rho(R)$ and $p(R)$ are expressed as
$\rho(R)/\rho_0 = 4({df}/{dR})^2 f^{-6}$ and
${p(R)}/{\rho_0} =  -4/3({d^2f}/{dR^2})f^{-5}$, 
where $f(R)={1}/{\sqrt{1+z(R)}}$.
These expressions of $\rho(R)$ and $p(R)$ 
 have quite interesting physical meanings.
The density is in proportion
to the square of the first derivative of $f$ with respect to the new
independent variable $R$, while the pressure is in proportion
 to the second derivative of $f$, that is, the curvature.
Therefore if the pressure is zero, $f(R)$ is the straight line
while the negative pressure corresponds to the positive curvature
of the curve $f(R)$ in $(R, f)$ plane. This is completely in contrast to
$(z, d_L(z))$ plane where as far as $\rho > 3p$, 
the curve $d_L(z)$ has the  positive curvature. Therefore in $(z, d_L(z))$
plane it is difficult to distinguish by eye if the pressure is negative 
or not. However in $(R, f)$ plane it is quite easy to
distinguish the sign of the curvature of the curve so that 
the negative pressure can be identified by eye.
To demonstrate this we show in Fig.1. the data from Supernova
Cosmology Project (\cite{perl99}) in $(R, f)$ plane.\footnote{
Similar analysis may be possible using data from High-Z Supernova
Search (\cite{schmidt98}). As an example  we use only the data from
Supernova Cosmology Project in this letter. This does not mean that
other data are not important.} 
Although the error bar is considerably large, a glance may show that
the curvature is positive. 

To analyze the data quantitatively, the first
way is to expand $f$ in a power series of $R$ as
$f(R)=1+\sum\limits_{n=1}^N f_nR^n$,
where $f_n$ are constants. We did this expansion
with $N=4$ for the data from Supernova
Cosmology Project (\cite{perl99}) and obtained 
\footnote{In the likelihood analysis, we used the inverse function
$R=R(z)$ by solving the quartic equation.}
$f_1=-0.5, ~~f_2=0.292^{+0.018}_{-0.015},~~ f_3=-0.257\pm0.034,$ and 
$~~f_4=0.046^{+0.060}_{-0.075}$.
Therefore the apparent positive curvature by eye is
confirmed quantitatively since $f_2$ is positive. \footnote{Note here
  that for the flat $\Lambda$ model with  $\Omega_M=0.3$ and $z=1$ which
  is roughly corresponding to the data of Supernova Cosmology Project 
(\cite{perl99}), the expansion of $f$ with  $N=4$ has an accuracy of
3.6\%. Here the accuracy means the relative error of $f$ computed from 
the expansion (3) to the given $f$ in Eq. (2).}

To know the convergence property of the expansion of $f$ in
a power series of $R$,  let us consider the flat $\Lambda$ model.
In this model,  $R$ is expressed as
\be
R=2\int^1_f{df}/{\sqrt{\Omega_M+(1-\Omega_M)f^6}},
\ee
where $\Omega_M$ is the present density parameter of the dark matter.
For $N=7$ the expansion is expressed as
\ba
f&=&1-\frac{1}{2}R+\frac{3}{8}(1-\Omega_M)R^2-\frac{5}{16}(1-\Omega_M)R^3
\nonumber\\
&&+\frac{5}{128}(1-\Omega_M)(7-3\Omega_M)R^4
-\frac{3}{256}(1-\Omega_M)(17\Omega_M-21)R^5\nonumber\\
&&+\frac{1}{1024}(1-\Omega_M)[8\Omega_M+168(1-\Omega_M)+
12\Omega_M(1-\Omega_M)+63(1-\Omega_M)^2]R^6\nonumber\\
&&-\frac{1}{14336}(1-\Omega_M)[8\Omega_M+1176(1-\Omega_M)
+132\Omega_M(1-\Omega_M)+1827(1-\Omega_M)^2]R^7.\ea
However in this expansion the sign of $f_n$ changes alternately
so that the convergence is extremely slow. The accuracy of the
expansion is only 16\% for $\Omega_M=0.3$ and $f=0.6$ corresponding to
$z=1.77$ and  $R=1.13$.  
While for low $z$, the accuracy is better. For example for
$\Omega_M=0.3$ and $f=0.7 ( R=0.79, z=1.04)$ the accuracy is 0.95\%.

The above expansion might  not be  accurate enough to analyze the
data from such as SNAP so that we need a Pade-like approximation in a
power series expansion of $f$. We adopt the following 
Pade-like approximation as
\be
f(R)= {1}/{\sqrt{u}},~~
u=1+\sum\limits_{n=1}^{N} u_nR^n.
\ee
For the  flat $\Lambda$ model, $R$ is related to $u$
as 
\be
R=\int^u_1{du}/{\sqrt{\Omega_Mu^3+(1-\Omega_M)}}.
\ee
Up to  $N=7$, $u$ is expanded as
\ba
u&=&1+R+\frac{3}{4}\Omega_MR^2+\frac{1}{2}\Omega_M R^3
+\frac{1}{16}\Omega_M(2+3\Omega_M)R^4\nonumber\\
&&+\frac{3}{16}\Omega_M^2R^5+\frac{1}{64}\Omega_M^2(4+3\Omega_M)R^6
+\frac{1}{112}\Omega_M^2(1+6\Omega_M)R^7.
\ea
All $u_n$ are positive definite so that the convergence
is very rapid. In reality for 
$\Omega_M=0.3$ and $f=0.6 (R=1.13, u=2.77)$ the relative 
error of $f$ is 0.059\%. In this expansion the error is the largest for
$\Omega_M=1$. However even in this case, the error is 0.2\% for
 $f=0.6 (R=0.8, u=2.77)$.
Therefore if accurate values of $R$ for various $z$ 
are obtained observationally,  we may determine
seven parameters $u_n (n=1,2, \ldots 7)$, which  may be enough to
express $f(R)$ in $ < 0.1\%$ accuracy. 

\section{ Equation of State}

Let us assume that accurate   $u_n$ for $n=1,\ldots, N$ are obtained
from the analysis of observational data. In this section we discuss how
to determine and confirm the equation of  state from  $u_n$.
Firstly $\rho(R)$ and $p(R)$ are expressed as
${\rho(R)}/{\rho_0} = ({du}/{dR})^2$ and
${p(R)}/{\rho_0} =- ({du}/{dR})^2  +{2u}/{3}({d^2u}/{dR^2})$. 
Secondly the square of the sound velocity $dp/d\rho$
is expressed as
\be
\frac{dp}{d\rho}=-\frac{2}{3}+\frac{u\frac{d^3u}{dR^3}}{3\frac{du}{dR}
\frac{d^2u}{dR^2}}.
\ee

\subsection{ One parameter Equation of State}

The  flat $\Lambda$ model has only one parameter, $\Omega_M$.
This parameter is equal to either $4u_2/3$ or $2u_3$ so that
the identity of $\Omega_M=4u_2/3=2u_3$ will be the consistency check 
of $\Lambda$ model (see also \cite{chiba98}).
{}From the data of Supernova
Cosmology Project (\cite{perl99}) we obtained 
$u_1=1, u_2=0.084^{+0.076}_{-0.063}, u_3=0.343^{+0.14}_{-0.13}$,
and $ u_4=0.360^{+0.32}_{-0.24}$.
Therefore 
$4u_2/3=0.112^{+0.101}_{-0.084}$ while
$2u_3=0.686^{+0.28}_{-0.26}$.
Note here that for the flat $\Lambda$ model
with  $\Omega_M=0.3$ and $z=1$ which is roughly corresponding to the
data of Supernova Cosmology Project (\cite{perl99}), the expansion 
of $u$ with $N=4$ has an accuracy of 0.19\%.
There is another null test of
the $\Lambda$ model for all $R$. {}From Eq.(5) we can derive
\ba
&&{du}/{dR}=\sqrt{\Omega_Mu^3+(1-\Omega_M)},\nonumber \\
&&{d^2u}/{dR^2}=\frac{3}{2}\Omega_M u^2,\nonumber \\
&&{d^3u}/{dR^3}=3\Omega_M u\sqrt{\Omega_Mu^3+(1-\Omega_M)}.
\ea
Using these expressions of derivatives, we can easily prove
$dp/d\rho=0$ for all $u$. Therefore null test of $dp/d\rho$ in Eq.(7)
observationally can confirm the $\Lambda$ model.
At present from the above values of $u_2$ and $u_3$, 
 we have $dp/d\rho$ at $u=1$ as ${dp}/{d\rho}=3.41^{+4.7}_{-5.2}$,
which means the present data is not accurate enough to
confirm the $\Lambda$ model.

\subsection{ Two parameter Equation of State}

The flat {\it w-cosmology} is an example of this class.
In {\it w-cosmology} the universe contains x-matter with
$p_X=w\rho_X$ where $w$ is a constant.
$R$ is expressed as
$R=\int^u_1{du}/{\sqrt{\Omega_Mu^3+(1-\Omega_M)u^{3(1+w)}}}$.
Then derivatives are given by
\ba
&&\frac{du}{dR}=\sqrt{\Omega_Mu^3+(1-\Omega_M)u^{3(1+w)}},\nonumber \\
&&\frac{d^2u}{dR^2}=\frac{3}{2}\Omega_M u^2+\frac{3}{2}
(1+w)(1-\Omega_M)u^{3w+2},\nonumber \\
&&\frac{d^3u}{dR^3}=[3\Omega_M u+\frac{3}{2}
(1+w)(3w+2)(1-\Omega_M)u^{3w+1}]\sqrt{\Omega_Mu^3+(1-\Omega_M)u^{3(1+w)}}.
\ea 
$u_2$ and $u_3$ are given by
\be
u_2=\frac{3\Omega_M}{4}+\frac{3(1+w)(1-\Omega_M)}{4},~~
u_3=\frac{\Omega_M}{2}+\frac{(1+w)(3w+2)(1-\Omega_M)}{4}.
\ee
 $\Omega_M$ and $w$ are determined from $u_2$ and $u_3$ as
\be
\Omega_M=1-\frac{(4u_2-3)^2}{12u_3-6-5(4u_2-3)},
~~w=\frac{4u_3-2}{4u_2-3}-\frac{5}{3}.
\ee
Then 
${dp}/{d\rho}$ ($={w(1+w)(1-\Omega_M)u^{2+3w}}/({\Omega_M u^2+(1+w)
(1-\Omega_M)u^{2+3w}}))$  should agree with  $dp/d\rho$ in Eq. (7)
so that we can confirm or refute {\it w-cosmology}. 

\subsection{General Equation of State}

Let us now consider the general case in which $w$ is a function of 
$u$ in  {\it w-cosmology}. $R$ is expressed as
\be
R=\int^u_1\frac{du}
{\sqrt{\Omega_Mu^3+(1-\Omega_M)\exp{(q(u))} }},~~
q(u)=\int^u_1\frac{3(1+w(u))}{u}du.
\ee 
Then derivatives are given by
\ba
&&\frac{du}{dR}=\sqrt{\Omega_Mu^3+(1-\Omega_M)\exp{(q(u))} },\nonumber \\
&&\frac{d^2u}{dR^2}=\frac{3}{2}\Omega_M u^2
+\frac{3}{2}(1-\Omega_M)\frac{(1+w(u))}{u}\exp{(q(u))},\nonumber \\
&&\frac{d^3u}{dR^3}=
[3\Omega_M u+\frac{3}{2}
(\frac{(1+w(u))}{u^2}(3w(u)+2)+\frac{dw}{udu})(1-\Omega_M)\exp{(q(u))}]
\sqrt{\Omega_Mu^3+(1-\Omega_M)\exp{(q(u))}}.\nonumber
\ea  
$u_2$ and $u_3$ are given by
\be
u_2=\frac{3\Omega_M}{4}+\frac{3(1+w_0)(1-\Omega_M)}{4},~~
u_3=\frac{\Omega_M}{2}+\frac{(1-\Omega_M)[(1+w_0)(3w_0+2)+w_1]}{4},
\ee
where $w_0=w(1)$ and $w_1=dw/du(u=1)$.
Now we have only two equations for three unknown constants 
$\Omega_M$, $w_0$ and $w_1$. To resolve this one may use the expression
for $u_4$. However in the expression of  $u_4$, a new unknown
constant $w_2=d^2w/du^2(u=1)$ appears so that we have to make
the closure. One way is to determine $\Omega_M$  from another data
 such as $d_L(z)$ for $z>3$  by NGST(Next Generation Space Telescope
 (\cite{eft99})). In this case $w_0$ and $w_1 \ldots$ are determined
as
\be
w_0=\frac{4u_2-3\Omega_M}{3(1-\Omega_M)}-1,~~
w_1=4u_3-2\Omega_M-(1+w_0)(3w_0+2),~~
w_2=\ldots 
\ee
The other is to assume $w_2=0$. Then
$\Omega_M$,$w_0$ and $w_1$ are determined as a solution
to simultaneous  non-linear  equations as
\be
u_2=u_2(\Omega_M,w_0,w_1),~~
u_3=u_3(\Omega_M,w_0,w_1),~~
u_4=u_4(\Omega_M,w_0,w_1).
\ee
In both cases $dp/d\rho$ is given by
\be
\frac{dp}{d\rho}=\frac{w(u)(1+w(u))+\frac{u}{3}w_1(1-\Omega_M)\exp{(q(u))}}
{\Omega_M u^2+(1-\Omega_M)\frac{1+w(u)}{u}\exp{(q(u))}},
\ee   
where $w(u)=w_0+w_1(u-1)$.
As before this $dp/d\rho$ should agree with $dp/d\rho$ in Eq. (7),
which is used to confirm or refute each  model.
In general, if $u_n$ are determined up to $N$,
 the equation of state with 
$ w=w_0+\sum\limits_{n=1}^{N-1}w_n(u-1)^n$
can be determined in principle.

\section{Discussion}

Now if x-matter consists of the scalar field $\phi$ with
the potential $V(\phi)$, they are related to $\rho_X$ and $p_X$
as
\ba 
\left(\frac{d\phi}{dt}\right)^2&=& \rho_X+p_X,\nonumber \\
V(\phi)&=&\frac{1}{2}(\rho_X-p_X). \nonumber
\ea
Using $\rho(R)$ and $p(R)$, we have
\ba
&&\phi-\phi_0=\frac{1}{\sqrt{8\pi
G}}\int_0^R\sqrt{-3\Omega_Mu+\frac{2}{u}\frac{d^2u}{dR^2}}dR\equiv g(R),\\
&&V(\phi)=\frac{3H_0^2}{\sqrt{16\pi G}}\left(2(\frac{du}{dR})^2-\Omega_Mu^3-
\frac{2u}{3}\frac{d^2u}{dR^2}\right)
\equiv h(R),
\ea
where $\phi_0$ is the present value of the scalar field . From Eq. (17) we
have
$R=g^{-1}(\phi-\phi_0).$  
Then the potential is expressed as
$V(\phi)=h(g^{-1}(\phi-\phi_0))$.

 In section 3 we assumed that  accurate   $u_n$ for $n=1,\ldots, N$
are obtained. We here show an example of the determination of $u_n$
for $n=1,\ldots, 7$. We adopt the redshifts of 38 data with $z > 0.17$
from (\cite{perl99}). We also adopt the relative error of $R$ for each
data as $X_i$. Now let us assume that our universe obeys the $\Lambda$
model with $\Omega_M=0.3$. Then we know the theoretical value of $R_i^t$
for each $z_i$. To simulate the real observation, we set
$R_i=R_i^t(1\pm SX_i)$ where S is a scale factor. 
Let us assume that an accurate observation gives us 
 38 luminosity distances with $\sim 0.1\%$ accuracy  so that
the scale factor $S$ is chosen to make the relative error of $R$ for
38 data be $\sim 0.1\%$.\footnote{This might be possible if
the statistical error approaches the systematic error in, for
example, the SNAP project. } We performed the likelihood analysis
for this simulated data and obtained $u_2=0.2240, u_3=0.1505,
u_4=0.0585$, $u_5=0.0187, u_6=0.0095$ and $ u_7=0.0053$ with $\chi^2=36.16$
for 38-7 d.o.f. while theoretical values are  $u_2=0.225, u_3=0.15,
u_4=0.054375$, $u_5=0.016875, u_6=0.0068906$ and $ u_7=0.00225$.
>From this $\Omega_M=4/3u_2=0.29866$ or  $\Omega_M=2u_3=0.3010$ is
obtained while $dp/d\rho=0.0052$ at $u=1$. Assuming the more
general equation of state with $w_2=w_3\ldots=w_ 7=0$,
we have $\Omega_M= 0.31095,  w_0= -1.01783$ and $w_1=-0.04768$.
This suggests that we may confirm the $\Lambda$ model if such accurate 
luminosity distances are available.
We show in Fig. 2 the simulated data and the results of the likelihood
analysis. Note that  error bars are extended by a factor 100.
 The theoretical curve (the dashed line) and the observational
curve (the solid line) are almost indistinguishable.

\acknowledgments{
  This work was supported in part by
Grant-in-Aid of Scientific Research of the Ministry of Education,
Culture, and Sports, No.11640274 and 09NP0801.}

\clearpage

\begin{figure}
\centerline{\epsfysize=6.5in \epsfbox{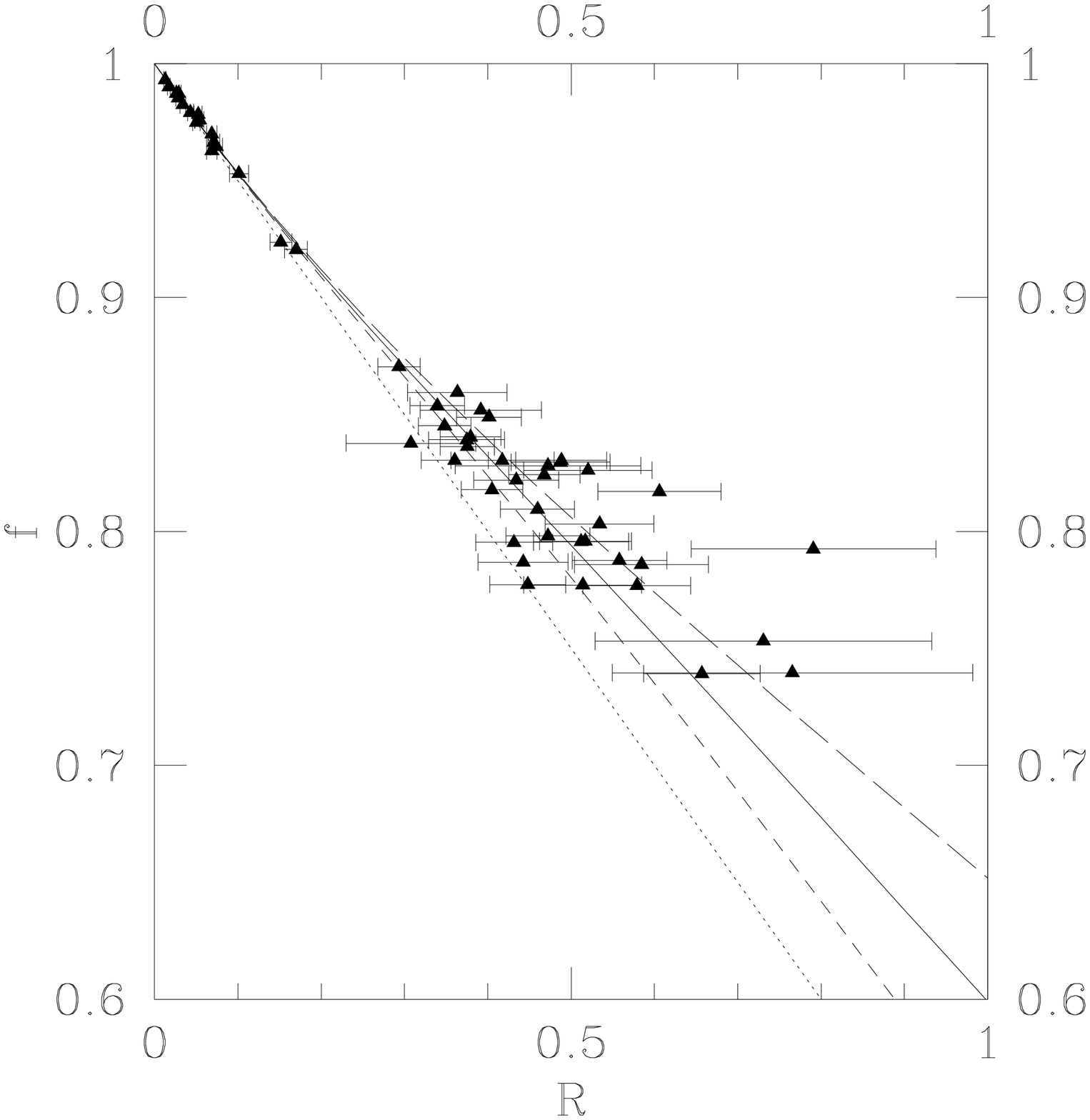}}
\figcaption[scp.ps]{54 TypeIa supernovae data from Supernova
 Cosmology Project (Perlmutter \etal 1999) in $(R, f)$ plane. 
 $f=1/\sqrt{1+z}$ and $R$ is a new independent variable defined by 
 $R\equiv (H_0/c)(d_L(z)/(1+z))$, where $H_0$ and $d_L(z)$ are 
 the present Hubble parameter and the luminosity distance,
 respectively.   The  dotted line shows the flat dust case $f=1-0.5R$.
 The solid line is given by the likelihood analysis;
 $f=1/\sqrt{u}, u=1+R+0.084R^2+0.343R^3+0.360R^4$  with $\chi^2=47.72$
 for 54-4 d.o.f. The long and short dashed lines correspond to 
 $\pm 1\sigma$ values of $u_2,u_3$ and $u_4$.}
\end{figure}

\begin{figure}
\centerline{\epsfysize=6.5in \epsfbox{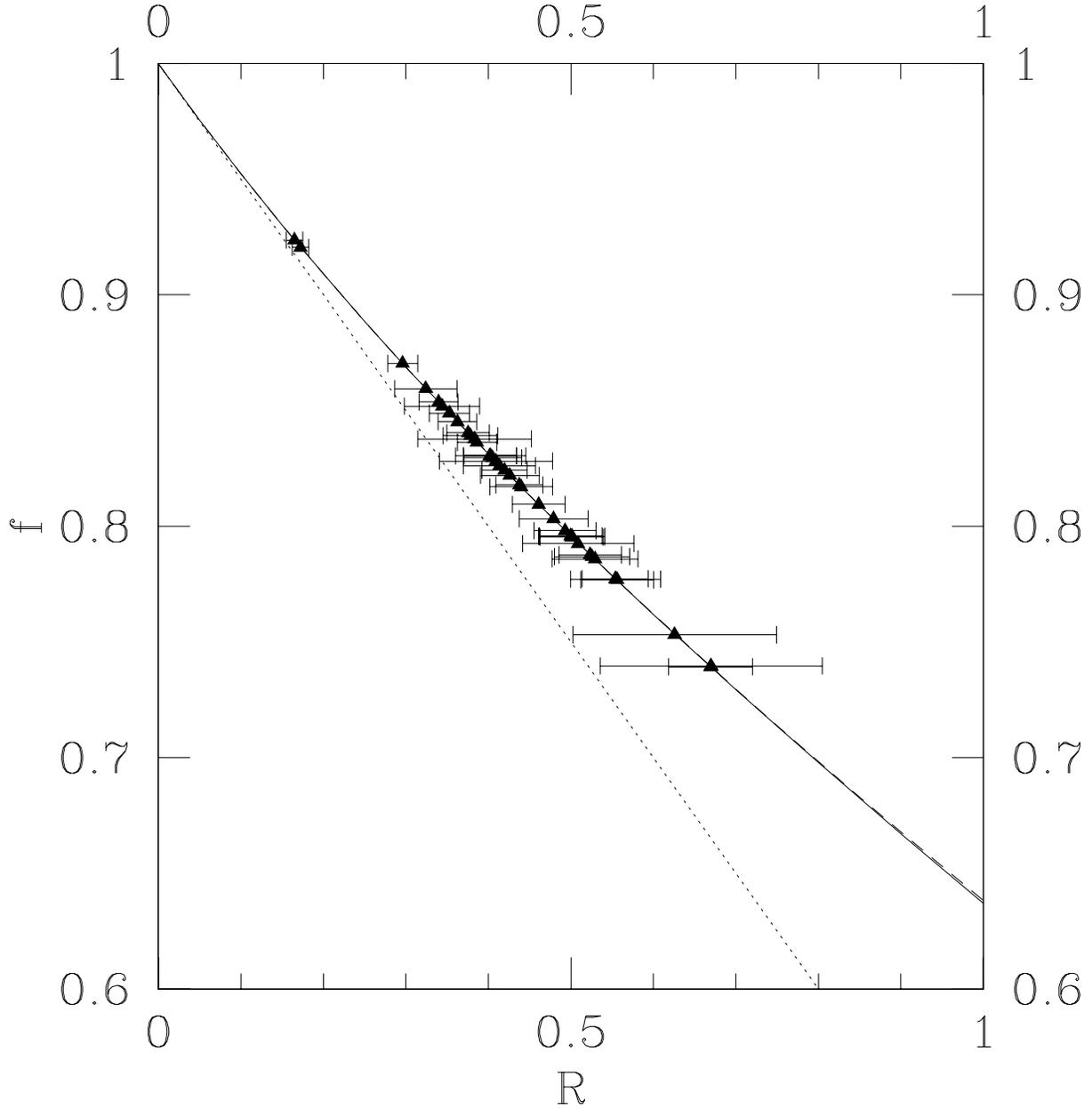}}
\figcaption[test.ps]{ 38 Simulated TypeIa supernovae data 
 in $(R, f)$ plane. See text how this data is made. The  dotted line
 shows the flat dust case $f=1-0.5R$. The solid line is the result of 
 likelihood analysis of
 $ f=1/\sqrt{1+R+0.2240R^2+0.1505R^3+
 0.0585R^4+0.0187R^5+0.0095R^6+0.0053R^7 }$ with  $\chi^2=36.16$
  for 38-7 d.o.f.. The dashed line is the theoretical curve of 
 $ f=1/\sqrt{1+R+0.225R^2+0.15R^3+
 0.054375R^4+0.016875R^5+0.0068906R^6+0.00225R^7 }$.
 Note that  error bars are extended by a factor 100.}
\end{figure}

\end{document}